\documentclass{ptapap}
\newcommand{\bz}{$\langle B_z \rangle$}
\author{Barnali Das}[NCRA]
\author{Poonam Chandra}[NCRA]
\author{Gregg A. Wade}[RMC]
\author{Matt E. Shultz}[UD]
\author{James Sikora}[QU]
\affil[NCRA]{National Centre for Radio Astrophysics, Tata Institute of Fundamental Research,  Pune University Campus, Pune-411007, India}
\affil[RMC]{Department of Physics, Royal Military College of Canada, PO Box 17000, Station Forces, Kingston, ON K7K 7B4, Canada}
\affil[UD]{Annie Jump Cannon fellow, Department of Physics and Astronomy, University of Delaware, 217 Sharp Lab, Newark, DE 19716, USA}
\affil[QU]{Queen's University, Ontario, Canada}

\title{Probing the magnetospheres of hot magnetic stars using ECME}

\begin{document}

\maketitle

\begin{abstract}

We discuss Electron Cyclotron Maser Emission (ECME), observed in the form of highly circularly polarized pulses, from a few hot magnetic stars. This emission is one of the manifestations of stellar wind-magnetic field interaction. With the Giant Metrewave Radio Telescope (GMRT), we have observed ECME from four magnetic B/A type stars. Currently, we are trying to understand certain properties of the ECME pulses and their dependences on the magnetospheric plasma. Here we briefly review all those works which have used ECME observed from hot magnetic stars to infer some physical properties of the host stars. We finally discuss how this phenomenon can further be exploited to probe the stellar magnetosphere.

 

\end{abstract}

\section{Introduction}
\begin{figure}
\includegraphics[width=0.7\textwidth]{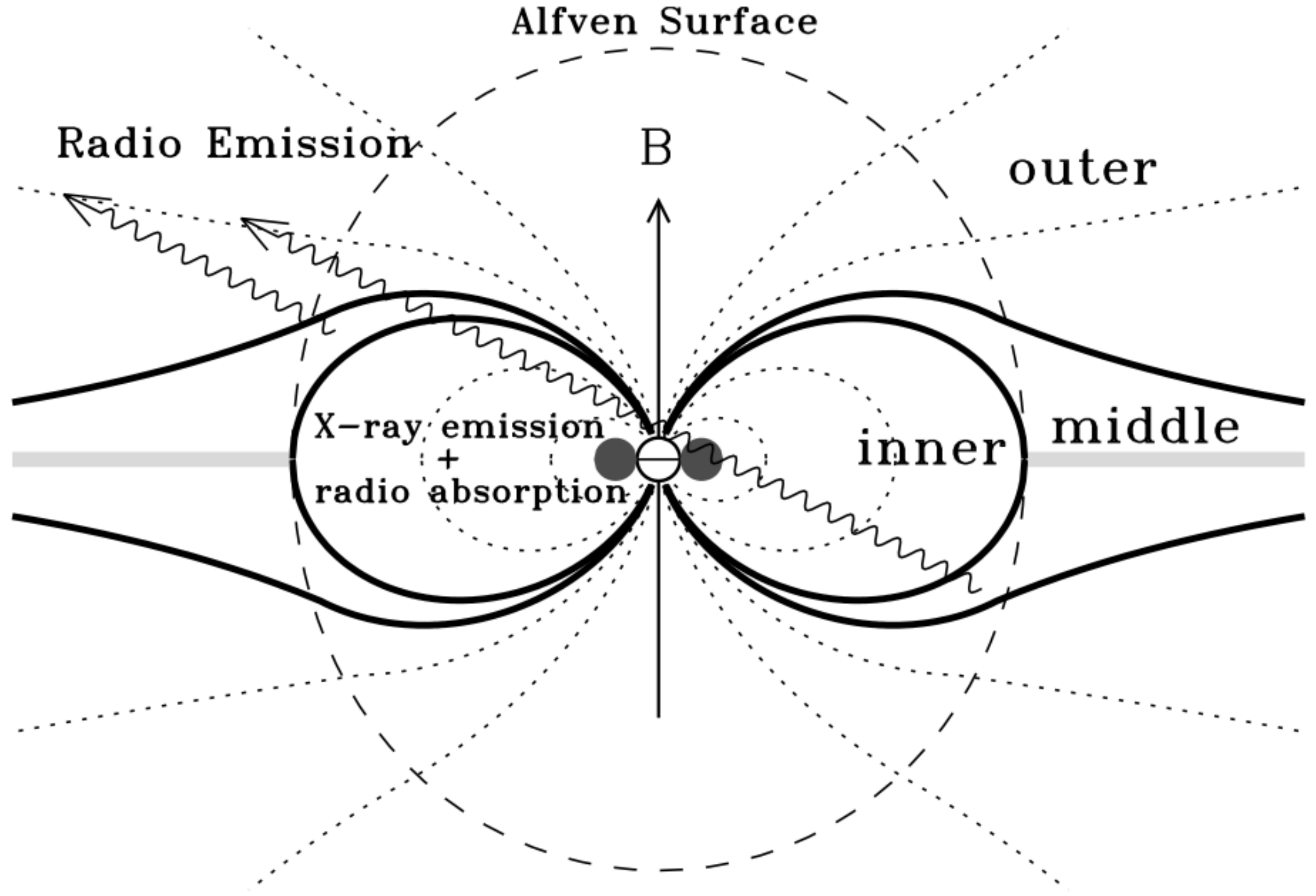}
\caption{The magnetosphere of a hot magnetic star with an axisymmetric dipolar magnetic field. This figure has been taken from \citet{trigilio2004}. The central circle represents the star and the vertical arrow represents the dipole axis. The magnetic field lines remain closed upto a certain distance (called the Alfv\'en radius) within which the magnetic energy dominates over the wind kinetic energy, this part of the magnetosphere is called the inner magnetosphere. Far away from the star, the wind dominates over the magnetic field and the field lines follow nearly radial topology. This part of the magnetosphere is named as outer magnetosphere. In between the two, the field lines are drawn into a current sheet which acts as the site for particle acceleration. This transition region is called the middle magnetosphere. The accelerated electrons propagate back towards the stellar surface following the field lines and emit gyrosynchrotron emission and in some cases ECME as well.}
\label{fig:3d_mag}
\end{figure}

Hot magnetic stars possess strong, ordered, mostly dipolar magnetic fields, and are expected to produce radio emission via gyrosynchrotron emission or free-free emission. A small number of them are also found to produce Electron Cyclotron Maser Emission (ECME) which is coherent in nature. The emission is believed to arise in the middle magnetosphere \citep{leto2016}, which is the transition region between inner (magnetic field dominates over the wind kinetic energy) and the outer magnetosphere \citep[wind kinetic energy dominates over the magnetic field,][]{trigilio2004}. The three regions of the magnetosphere are shown in Fig. \ref{fig:3d_mag} \citep[taken from][]{trigilio2004}. 
The emission is highly circularly polarized and highly directed; for a mildly relativistic electron distribituion, this direction is almost perpendicular to the magnetic field \citep{melrose1982}. The latter property is consistent with the fact that the ECME pulses have been observed near the rotational phase where the longitudinal magnetic field is zero, i.e. when the magnetic field axis lies in the plane of the sky (assuming a dipolar magnetic field).

The first hot magnetic star, from which ECME was discovered, is CU\,Vir \citep{trigilio2000}. After this discovery, there was not a second one until 2015, when \citeauthor{chandra2015} speculated HD\,133880 to host similar emission after they observed an order of magnitude enhancement in the flux density at certain rotational phases. We confirmed it \citep{das2018} by observing the star at 610 MHz for one full rotation cycle using the Giant Metrewave Radio Telescope (GMRT). 

One of the biggest questions about ECME from hot magnetic stars is why this emission is not seen from most of the population. In order to be able to answer this question, we have observed a sample of hot magnetic stars with the GMRT at 550--900 MHz, near the rotational phases corresponding to the nulls of the longitudinal magnetic field (\bz). Besides performing this survey, we also study certain ECME properties which have the potential to become tools to estimate properties of the magnetospheric plasma. Here, we briefly review the already existing methods to extract information about the host star from the ECME pulses, and also present some new directions in which this emission can become useful to probe the stellar magnetosphere.

This proceeding paper is structured as follows: in the next section (\S \ref{gmrt_survey}), we present the result of the GMRT survey. Following this section, we review how ECME has been used to study the host star and then we present how we can further use this emission to infer plasma properties in the stellar magnetosphere (\S \ref{ecme_tool}). We will finally discuss various results and future work in \S \ref{sec:disc}.


\section{Results from the GMRT survey}\label{gmrt_survey}
At the time of starting this survey, CU\,Vir and HD\,133880 were the only known hot magnetic stars producing ECME. The fact that within a span of around 17 years, only two stars were detected, gives the impression that for some unknown reason, ECME is extremely rare. However, we realized that there was not any systematic search for this kind of emission, and in fact, there exist very few observations of hot magnetic stars below 1.4 GHz. This motivated us to start a low radio frequency survey with the upgraded GMRT (uGMRT) to search for ECME. We choose band 4 (550-900 MHz) as the observing band. 


So far we have made two discoveries: HD\,142990 and HD\,35298. Among them, HD\,142990 was independently reported to be a tentative host of ECME by \citet{lenc2018}. We have also found a tantative candidate Ap star which will be further observed for confirmation. With the two confirmed stars, the current number of confirmed hot magnetic stars capable of producing ECME is five: CU\,Vir \citep{trigilio2000}, HD\,133880 \citep{chandra2015,das2018}, HD\,142990 \citep{lenc2018,das2019a}, HD\,35298 \citep{das2019b} and HD\,142301 \citep{leto2019}.

In the next two subsections, we describe various characteristics of ECME observed from HD\,142990 and HD\,35298.

\subsection{HD 142990}
\begin{figure}
\includegraphics[width=0.9\textwidth]{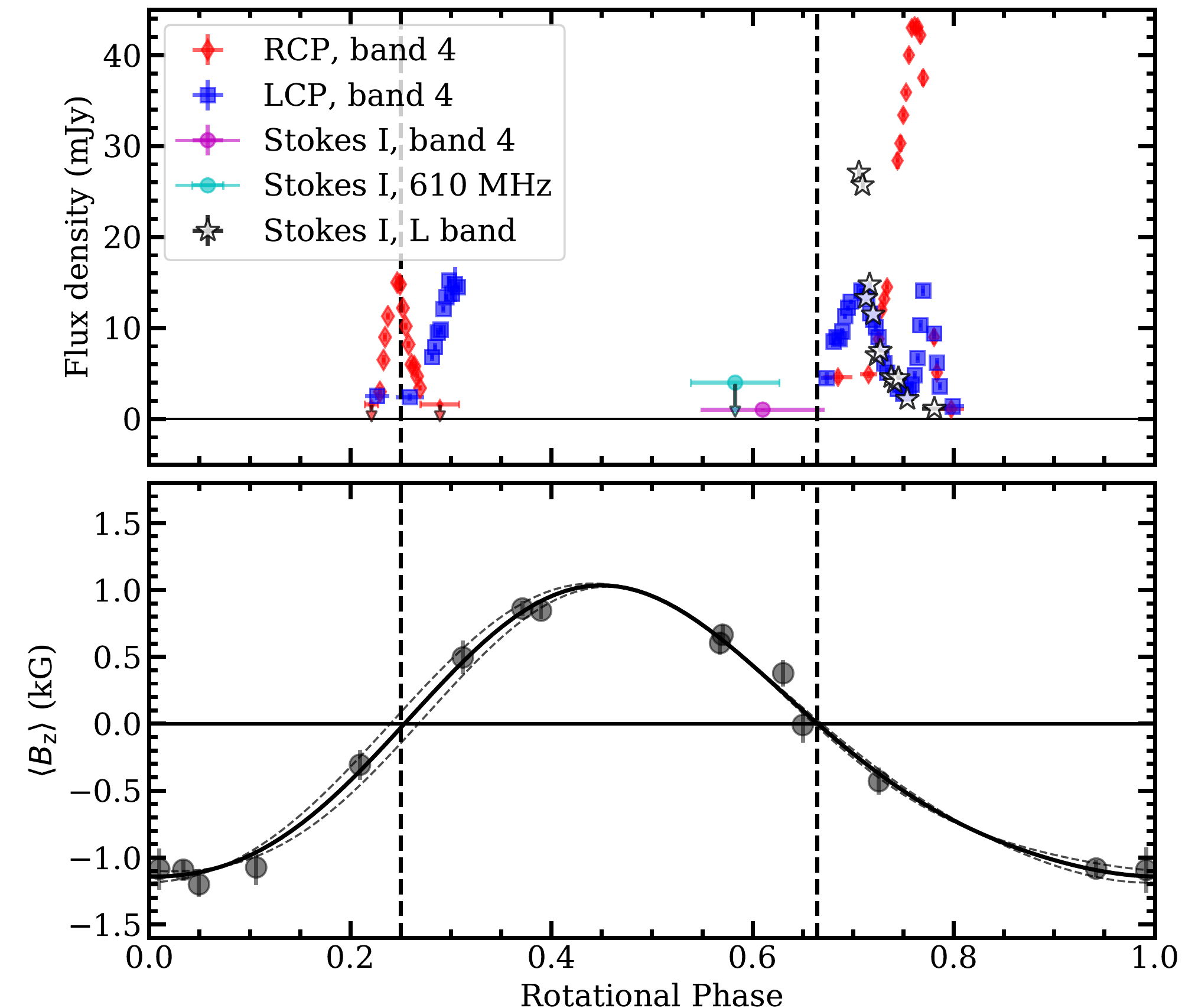}
\caption{The ECME pulses observed from HD\,142990 (upper panel). The longitudinal magnetic field of the star is shown in the lower panel. This figure is taken from \citet{das2019a}.}
\label{fig:hd142990}
\end{figure}

HD\,142990 is a Bp type star with a surface magnetic field of strength 4.7 kG \citep{shultz2019}. \citet{lenc2018} observed highly cirularly polarized emission at 200 MHz from HD\,142990 using the Murchison Widefield Array (MWA) from which they speculated it to be a probable host of ECME. We independently observed this star as part of our ongoing survey near the rotational phases of its two magnetic nulls. The result of our observation is shown in Fig. \ref{fig:hd142990} \citep{das2019a}. We detected highly circularly polarized pulses near both magnetic nulls which confirmed the ECME origin of the pulses. In addition, we also detected the signature of magneto-ionic mode transition which can be confirmed only by future observation at frequencies higher and lower than our current observing frequency (550-900 MHz). HD\,142990 is the most massive hot magnetic star \citep[5.7 $M_\odot$,][]{shultz2019} from which ECME has been observed.

\subsection{HD35298}
\begin{figure}
\includegraphics[width=0.9\textwidth]{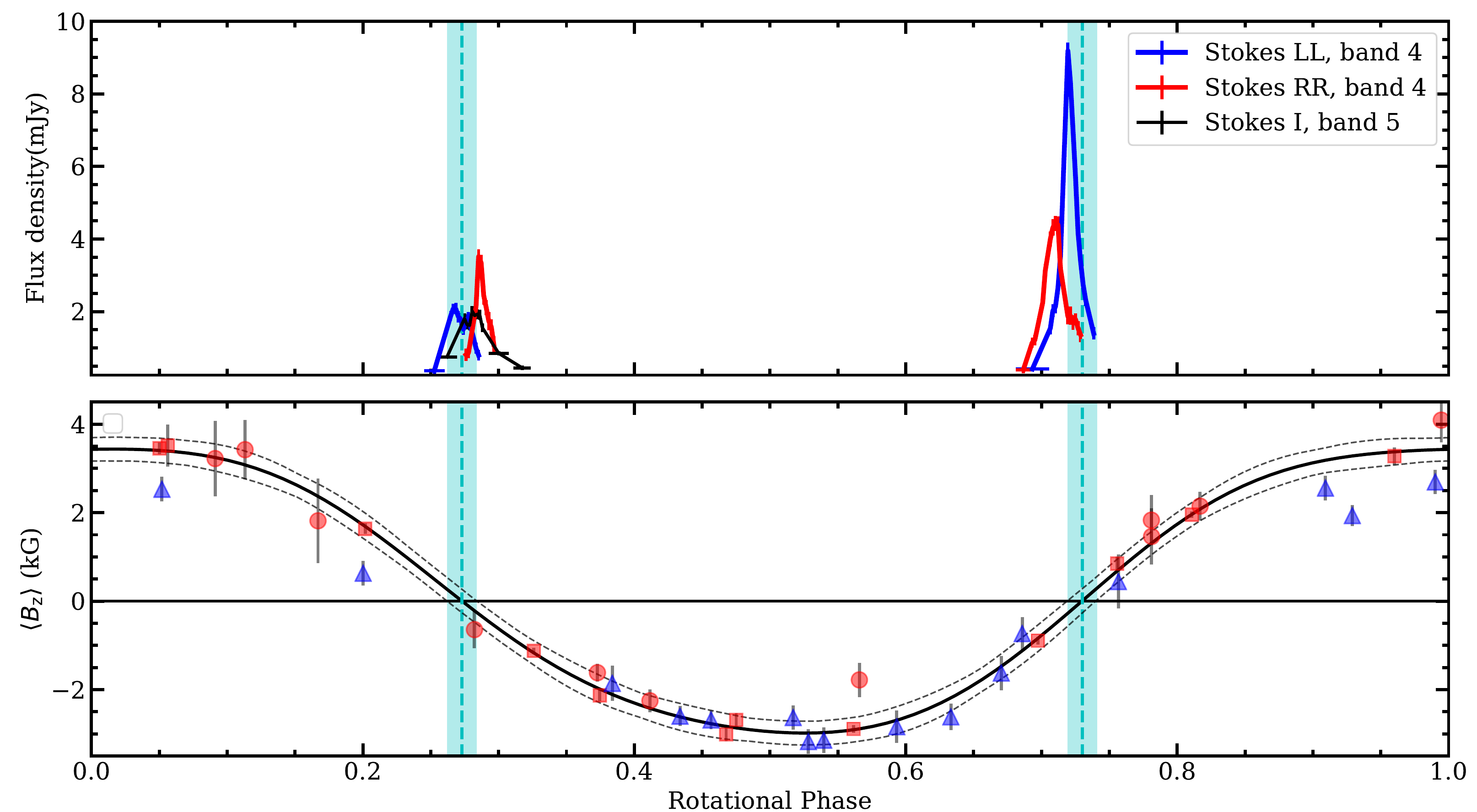}
\caption{The ECME pulses observed from HD\,35298 (upper panel). The longitudinal magnetic field of the star is shown in the lower panel. This figure is taken from \citet{das2019b}.}
\label{fig:hd35298}
\end{figure}

HD\,35298 is also a chemically peculiar B type star with a nearly dipolar magnetic field of polar strength 10.8 kG \citep{shultz2019}. The rotation period of the star is around 1.85 days \citep{shultz2018} which makes it the slowest host of ECME among the hot magnetic stars till now. The ECME pulses observed from this star are shown in Fig. \ref{fig:hd35298} \citep{das2019b}.

\section{ECME as a tool to probe the stellar magnetosphere}\label{ecme_tool}
Since its discovery in CU\,Vir \citep{trigilio2000}, the potential of ECME  to become a tool to constrain various stellar/magnetospheric properties has been reported in a number of studies. We briefly review past works aimed at extracting information from the observed ECME pulses in the next subsection. Following it, we present another property of ECME, characterization of which can give us valuable information about the inner magnetosphere. 

\subsection{Past work}
Because of the high directivity, ECME pulses have been used to estimate the change in stellar rotation period \citep[e.g.][]{trigilio2008}. The principal idea is that if a star is rotating with a constant period, the ECME pulses will always arrive at the same rotational phases. Using this property, \citet{trigilio2008} proposed a spin-down of CU\,Vir. However, \citet{pyper2013} reported that the period change indicated by radio data is inconsistent with that obtained from optical photometry and suggested that the radio emitting region is probably not in synchronous rotation with the photophere. Two alternate explannations to the shift of the radio pulses are that the emission region is itself drifting, or that the emission region is  unstable \citep{ravi2010}.

\begin{figure}
\includegraphics[width=0.8\textwidth]{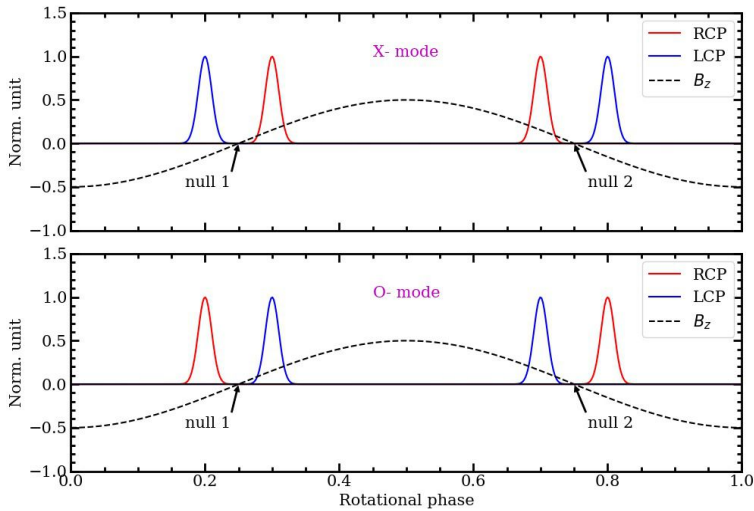}
\caption{A cartoon diagram illustrating the difference in the ECME lightcurves between X-mode (upper panel) and O-mode (lower panel).}
\label{fig:mock_x_O_mode_lightcurves}
\end{figure}

Another important information that we can extract about the stellar magnetosphere from ECME pulses is the magneto-ionic mode of emission. The mode of emission determines the circular polarization of the pulses. This is shown in the cartoon diagram Fig. \ref{fig:mock_x_O_mode_lightcurves}. If we detect ECME from both magnetic hemispheres and the longitudinal magnetic field $\langle B_\mathrm{z}\rangle$ of the star is known, we can infer the mode of emission \citep{leto2019}. This is useful as it helps us to constrain the plasma density at the site of emission. If the ratio between plasma frequency ($\nu_\mathrm{p}$) to the electron cyclotron frequency ($\nu_\mathrm{B}$) is less than 0.3--0.35, the mode of emission is extra-ordinary (X-) mode and above it, the mode will be ordinary (O-) mode. \citep{melrose1984,sharma1984}.

An interesting property of ECME is that the pulses at different frequencies arrive at slightly different rotational phases. This was attributed to propagation effect by \citet{trigilio2011}. Based on this model, \citet{lo2012} successfully reproduced the pulse arrival sequence of ECME at 20 cm and 13 cm. The salient features of this model are \citep{trigilio2011,lo2012}:

\begin{itemize}
\item ECME is emitted perpendicularly to the magnetic dipole axis as well as the local magnetic field line irrespective of the frequency of emission.
\item Density in the inner magnetosphere is constant.
\item There is only a single refraction at the boundary between middle and inner magnetospheres at the time of entering the latter. Refraction at the time of exiting the inner magnetosphere is neglected.
\end{itemize}


In the next subsection we show how we can use this simple model to derive a relation between pulse arrival time and frequency. 

\begin{figure}
\includegraphics[width=0.9\textwidth]{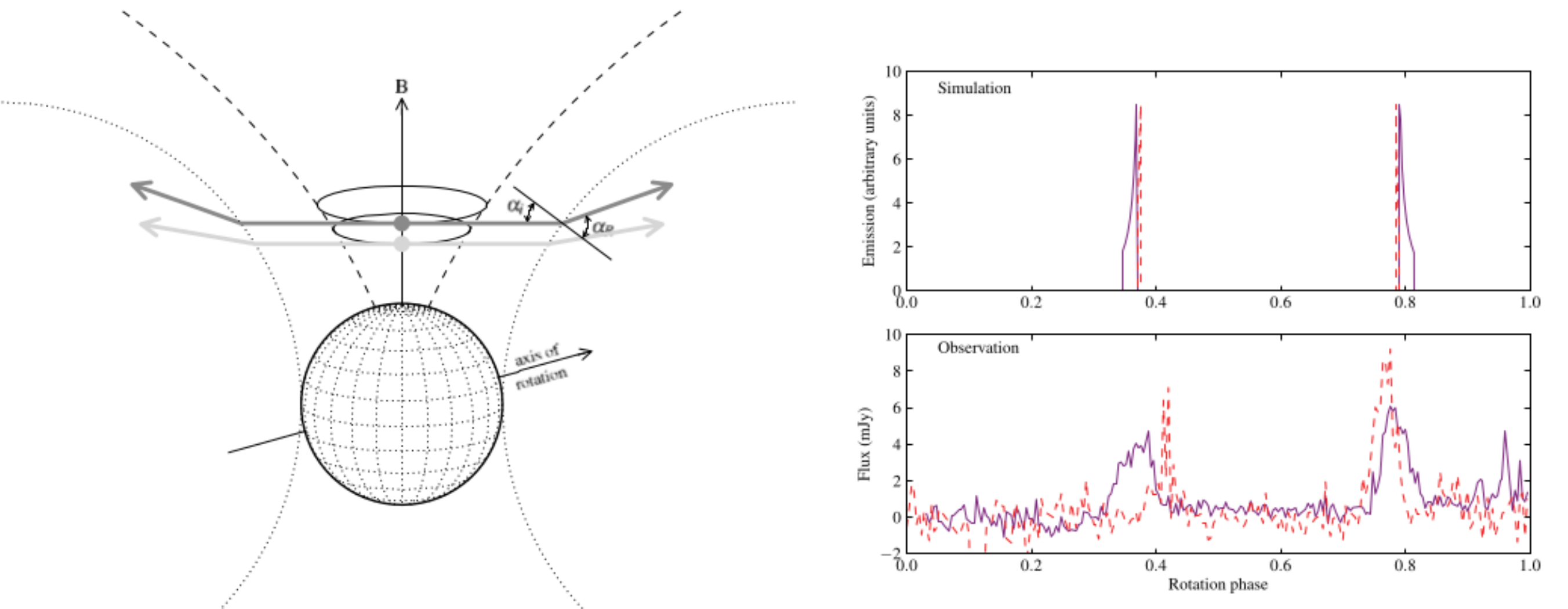}
\caption{\textit{Left}: The model proposed by \citet{trigilio2011} to explain the difference in arrival times of ECME pulses at different frequencies. ECME is emitted at $90^\circ$ to the dipole axis and also the local magnetic field line in the middle magnetosphere, at all frequencies. While entering the inner magnetosphere from the middle magnetopsphere, the ray suffers refraction at the boundary, which is different at different frequencies leading to different arrival times of the pulses at different frequencies. \textit{Right}: The lower panel shows the observed lightcurves of CU\,Vir \citep[by][]{lo2012} at 20 cm (solid line) and 13 cm (dotted line). The upper panel shows the simulated lightcurves at the two frequencies using the model of \citet{trigilio2011}. This figure has been taken from \citet{lo2012}.}
\label{fig:1}
\end{figure}

\subsection{Frequency dependence of pulse arrival time: a new way to estimate plasma density}

\begin{figure}
\includegraphics[width=0.6\textwidth]{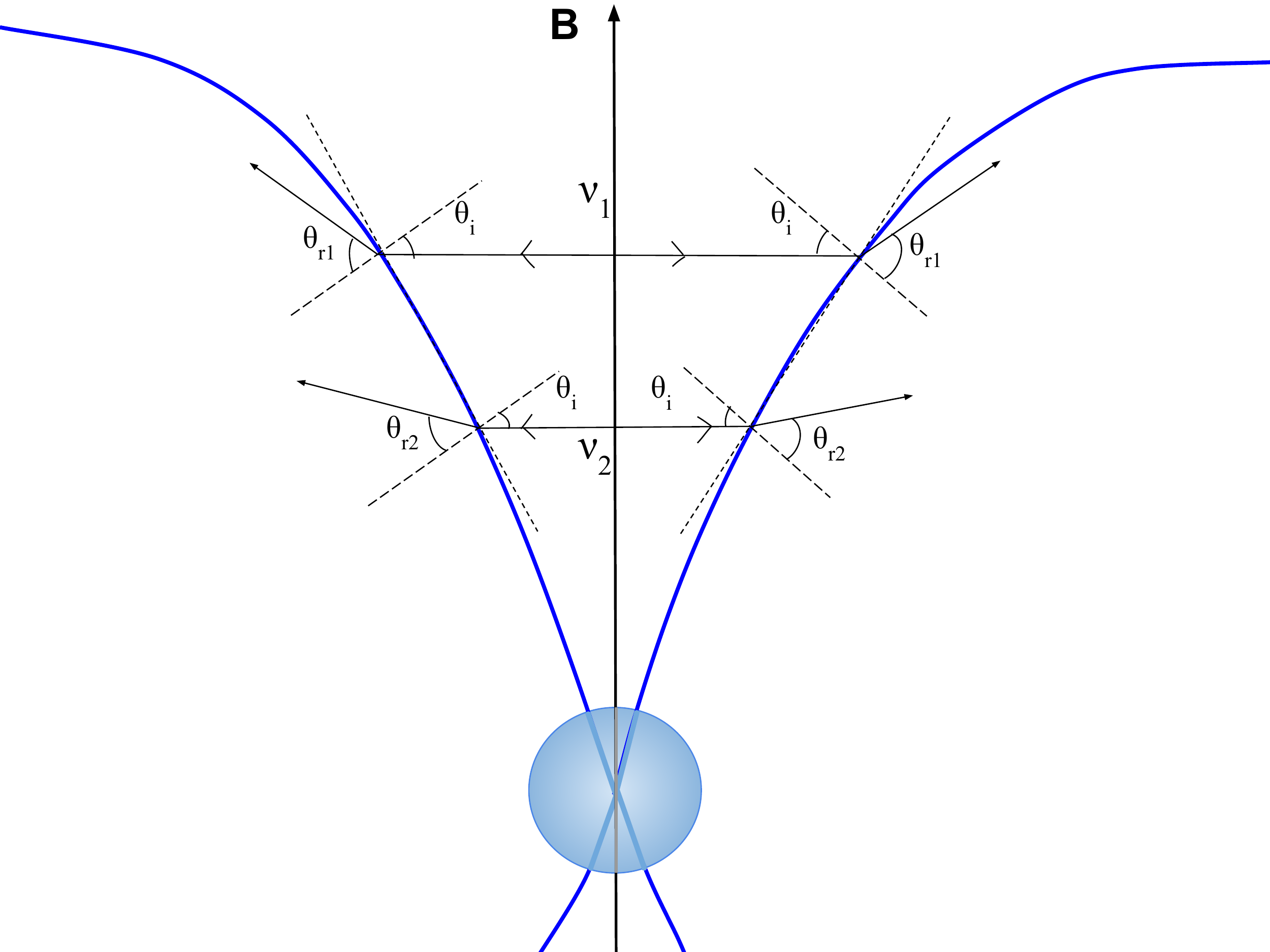}
\caption{The geometry of the stellar magnetosphere assumed while deriving Eq. \ref{eq:lag}. The blue shaded circle represents the star and the blue curved lines represent the boundary of the inner magnetosphere. The vertical arrow represents the magnetic dipole axis. ECME is emitted at a right angle to the magnetic field axis and is incident on the boundary between inner and middle magnetospheres at angle $\theta_i$. This angle is assumed to be same for both the frequencies $\nu_1$ and $\nu_2$, with $\nu_1<\nu_2$. The angle of refraction for the two frequencies are determined by the corresponding refractive indices $\mu_1$ and $\mu_2$; the angle of refractions are $\theta_{r1}$ and $\theta_{r2}$ respectively. Note that the deviations of the rays are exaggerated for clarity.}
\label{fig:model}
\end{figure}

The aim of this work is to quantify the dependence of pulse arrival time on frequency. 
We derive a mathematical relation between the difference in rotational phases of pulse arrival at two frequencies (i.e., lag, denoted by $\tau_{12}$) and the two frequencies $\nu_1$ and $\nu_2$ for the model of \citet{trigilio2011}. Assuming $\nu_\mathrm{p}/\nu_\mathrm{B}<<1$, we can write the refractive index as $\mu=\sqrt{1-\frac{A\nu_\mathrm{p}^2}{\nu^2}}$, where $A$ is a constant that depends on the magneto-ionic mode and harmonic number. At the time of entering the inner magnetosphere, the emission gets refracted and the angle of refraction ($\theta_r$) can be obtained using Snell's law. For an angle of incidence of $\theta_i$, the difference in angle of refractions at two frequencies is given by:

\begin{equation}
\theta_{r2}-\theta_{r1}=\sin^{-1}\left(\frac{\sin\theta_i}{\mu_2}\right)-\sin^{-1}\left(\frac{\sin\theta_i}{\mu_1}\right)
\end{equation}

The various angles used in this equation are shown in Fig. \ref{fig:model}. The corresponding lag is expected to be proportional to this difference. If we further assume that the lags are small, then to leading order, we get:

\begin{equation}
\tau_{21}\propto\theta_{r2}-\theta_{r1}=C\frac{\nu_2^2-\nu_1^2}{\nu_1^2\nu_2^2}=\tilde{C}(\lambda_1^2-\lambda_2^2)\label{eq:lag}
\end{equation}

The constant $C$ is a function of the plasma density in the inner magnetosphere and $\tilde{C}=C/c^2$, $c$ being the speed of light . Thus, if we have wideband observations (simultaneous, to avoid the complexity that can arise due to change in rotation period or/and a drifting/unstable emission region), we can obtain the lags and from the observed lags, we can estimate the inner magnetospheric density.

\section{Discussion}\label{sec:disc}
It is not known yet why the magnetospheres of some of the hot magnetic stars can produce ECME and why others cannot. We are trying to solve this problem by systematically observing a sample of magnetic B-type stars with known magnetic properties with the uGMRT near their magnetic nulls. We detected ECME from two of them. 

ECME pulses have a number of interesting properties which have the potential to become probes for the stellar magnetosphere. Here we demonstrate the potential of one such property which is the frequency dependence of pulse arrival time. This exploits the refractive index of the inner magnetosphere and hence constrains the density in the inner magnetosphere. Note that, for this method to give a reliable estimate, it is important to ensure that the data at different frequencies are near-simultaneous. 

Currently, our understanding about ECME from hot magnetic stars is highly limited. In addition to the fact that we do not know yet which subset of these stars can produce this type of emission, we also do not know the cut-off frequencies. Besides, ECME pulses are often found to be offset from their expected rotational phases, the reason for which is again unclear, although complex magnetic field topology is speculated to be at its root \citep[e.g.][]{leto2019, das2019a}. Both spectropolarimetric observations and radio observations over a wide frequency range will be needed to answer these questions.   


\acknowledgements{BD thanks Surajit Mondal for very fruitful discussions. PC  acknowledges support from the Department of Science and Technology via SwarnaJayanti Fellowship awards (DST/SJF/PSA-01/2014-15). GAW acknowledges Discovery Grant support from the Natural Sciences and Engineering Research Council (NSERC) of Canada. MES acknowledges support from the Annie Jump Cannon Fellowship, supported by the University of Delaware and endowed by the Mount Cuba Astronomical Observatory. We thank the staff of the GMRT that made these observations possible. 
The GMRT is run by the National Centre for Radio Astrophysics of the Tata Institute 
of Fundamental Research. This research has made use of NASA's Astrophysics Data System.}

\bibliographystyle{ptapap}
\bibliography{das}

\end{document}